\documentclass[fleqn,12pt,twoside]{article}
\usepackage[headings]{espcrc1}

\readRCS
$Id: espcrc1.tex,v 1.2 2004/02/24 11:22:11 spepping Exp $
\usepackage{graphicx}
\usepackage[figuresright]{rotating}

\newcommand{\AmS}{{\protect\the\textfont2
  A\kern-.1667em\lower.5ex\hbox{M}\kern-.125emS}}

\title{Isotopic effects in multifragmentation and the nuclear 
       equation of state}

\author{W. Trautmann\address[GSI]{Gesellschaft f{\"u}r Schwerionenforschung mbH,\\ 
        D-64291 Darmstadt, Germany} and the ALADIN and INDRA collaborations} 

\runtitle{Isotopic effects in multifragmentation and the nuclear 
       equation of state}
\runauthor{W. Trautmann}

\begin{document}

% typeset front matter
\maketitle

\begin{abstract}
Isotopic effects in spectator fragmentations following heavy-ion collisions at relativistic energies are investigated using data from recent exclusive experiments with SIS beams at GSI. Reactions of $^{12}$C on $^{112,124}$Sn at incident energies 
300 and 600 MeV per nucleon were studied with the INDRA multidetector while the 
fragmentation of stable $^{124}$Sn and radioactive $^{107}$Sn and $^{124}$La 
projectiles was studied with the ALADIN spectrometer. 

The global characteristics of the reactions are very similar. This includes the rise and fall of fragment production and deduced observables as, e.g., the breakup temperature obtained from double ratios of isotope yields. The mass distributions depend strongly on the neutron-to-proton ratio of the decaying system, as expected for a simultaneous statistical breakup. The ratios of light-isotope yields from neutron-rich and neutron-poor systems follow the law of isoscaling. The deduced scaling parameters decrease strongly with increasing centrality to values smaller than 50\% of those obtained for the peripheral event groups. This is not compensated by an equivalent rise of the breakup temperatures which suggests a reduction of the symmetry term required in a liquid-drop description of the fragments at freeze-out.
\end{abstract}

\section{INTRODUCTION}

Two themes are presently prevailing in the discussion of nuclear 
multifragmentation. One of them concerns the order of the nuclear 
liquid-gas phase transition. It should be of first order 
according to the range dependence of the nuclear forces but phenomena 
characteristic of second-order transitions are equally present in 
multifragmentation data~\cite{traut_inpc,sfienti_nn2006,topvolepja}.
The second class of questions is related to the nuclear equation of 
state and to its importance for astrophysical processes. Supernova simulations 
or neutron star models require inputs for the nuclear equation of state 
at extreme values of density and asymmetry for which the predictions 
differ widely \cite{lattprak,botv04}. Fragmentation reactions are of interest
here because they permit the production of nuclear systems 
with subnuclear densities and 
temperatures which, e.g., largely overlap with those expected for the 
explosion stages of core-collapse supernovae \cite{botv05}. 
Laboratory studies of the properties of nuclear matter in the hot 
environment, similar to the astrophysical situation, are thus becoming feasible, 
and an active search for suitable observables 
is presently underway. Of particular interest is the 
density-dependent strength of the symmetry term which is essential for the 
description
of neutron-rich objects up to the extremes encountered in neutron stars 
(refs.~\cite{bao02,greco02,yong_arx} and contributions to this conference).

Here, new results from two recent experiments performed at the 
GSI laboratory will 
be discussed with particular emphasis on the second class of topics.
The INDRA multidetector~\cite{pouthas} 
has been used to study isotopic effects in nearly
symmetric $^{124,129}$Xe on $^{112,124}$Sn reactions at incident energies up to 150 
MeV per nucleon and spectator fragmentations in $^{12}$C on $^{112,124}$Sn 
at incident energies 300 and 600 MeV per nucleon. 
In a very recent experiment with the ALADIN spectrometer, the possibility 
of using secondary beams for reaction studies at relativistic energies 
has been explored. Beams of $^{107}$Sn, $^{124}$Sn, $^{124}$La, and $^{197}$Au 
as well as Sn and Au targets
were used to investigate the mass and isospin dependence of projectile 
fragmentation at 600 MeV per nucleon. The neutron-poor radioactive 
projectiles $^{107}$Sn and $^{124}$La were produced at the Fragment Separator FRS
by fragmentation of a primary beam of $^{142}$Nd and delivered to the ALADIN 
experiment.

The data analysis performed so far includes isotopic effects in neutron and 
fragment production, isoscaling and its relation to the properties of hot 
fragments at the low-density freeze-out, 
flow in mass-symmetric collisions at intermediate 
energy, transparency in central collisions at intermediate energies as 
obtained from isospin tracing, as well as size fluctuations of the heaviest 
fragment of the partition and their relation to the order of the phase 
transition \cite{sfienti_nn2006}. In the following, emphasis will be given
to phenomena encountered in spectator fragmentation which is believed to occur 
at low density in the region of liquid-gas coexistence of the nuclear-matter 
phase diagram.

\section{THE PHASE DIAGRAM}

A phase diagram of nuclear matter as it is of interest in multifragmentation has
been discussed some years ago at a previous conference of this series 
(Fig. 1, bottom left, from ref.~\cite{nn2000}). It is partly
schematic, with the boundary of coexistence (CE) and the adiabatic spinodal (AS)
adapted from ref.~\cite{muell95}, and partly theoretical, with critical points as
derived for nuclear matter \cite{muell95,jaqa83} and for a $^{16}$O nucleus 
in a container \cite{schnack97}. 
The three experimental points representing the freeze-out
conditions in spectator fragmentation were obtained by correlating results from
temperature measurements and from interferometry performed in ALADIN experiments
\cite{xi97,fritz99}. 

In more recent years, the established systematics of
freeze-out temperatures \cite{nato02} and alternative indications for the expansion 
prior to freeze-out \cite{nn2003,viola04} have confirmed this general picture.
Freeze-out occurs in the coexistence zone below the critical temperature and at
densities equivalent to a fraction of the normal nuclear density.
New evidence has also been obtained for the equilibrium nature of the 
freeze-out configurations. 
For example, 
the systematics of fragment 
multiplicity versus excitation energy compiled by Tamain \cite{tamain06} 
shows that multifragmentation is a universal phenomenon driven by the thermal 
excitation of the system, fairly independent of the type of collision through 
which it is imparted. The achievement of equilibrium is important for the
astrophysical implications as the time scales encountered there are slow,
corresponding to equilibrium situations on the nuclear scale.

\begin{figure}[htb]
\centering
\begin{minipage}[c]{.49\textwidth}
   \centerline{\includegraphics[width=0.90\textwidth]{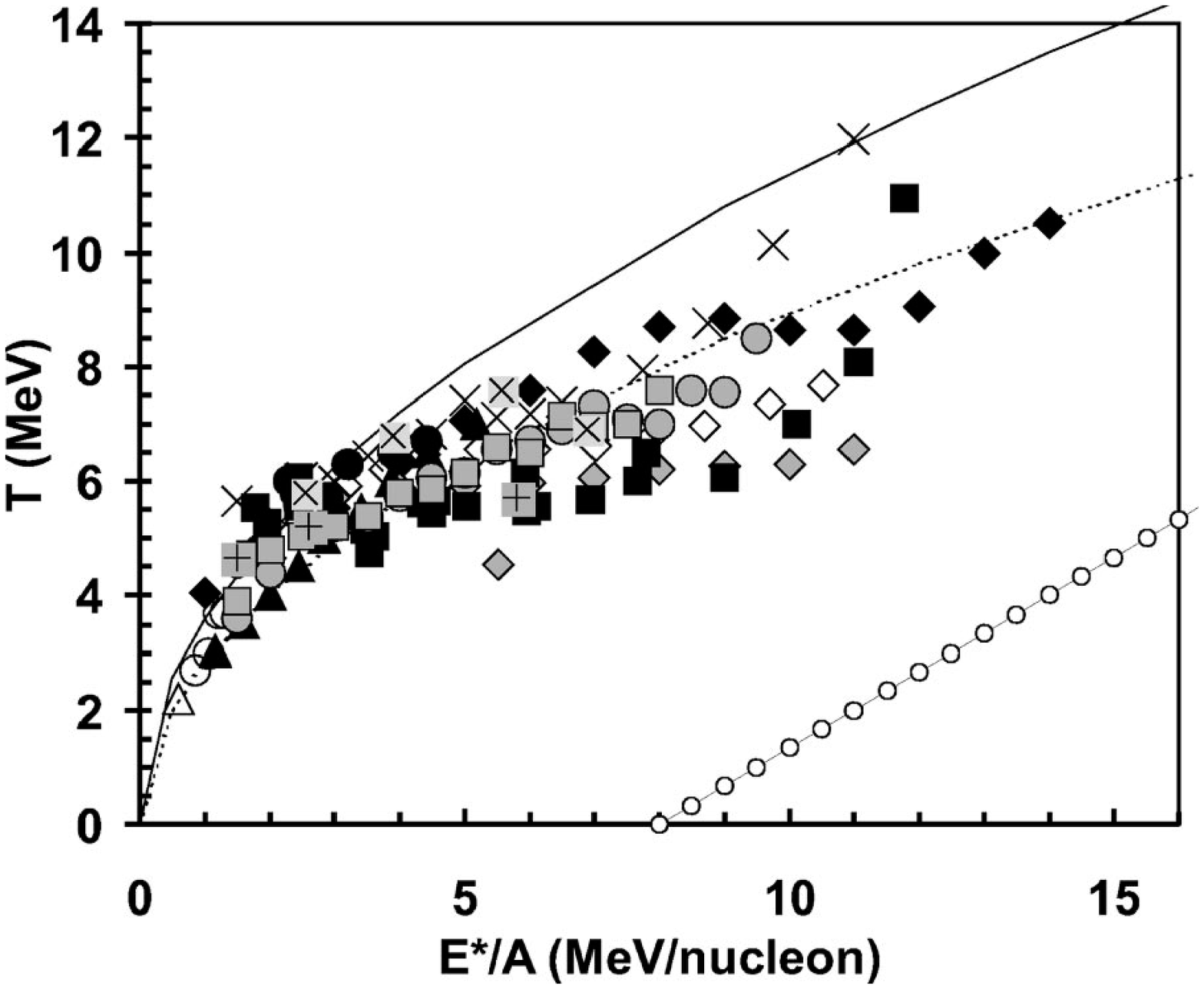}}
\vspace{9mm}
   \includegraphics[width=0.90\textwidth]{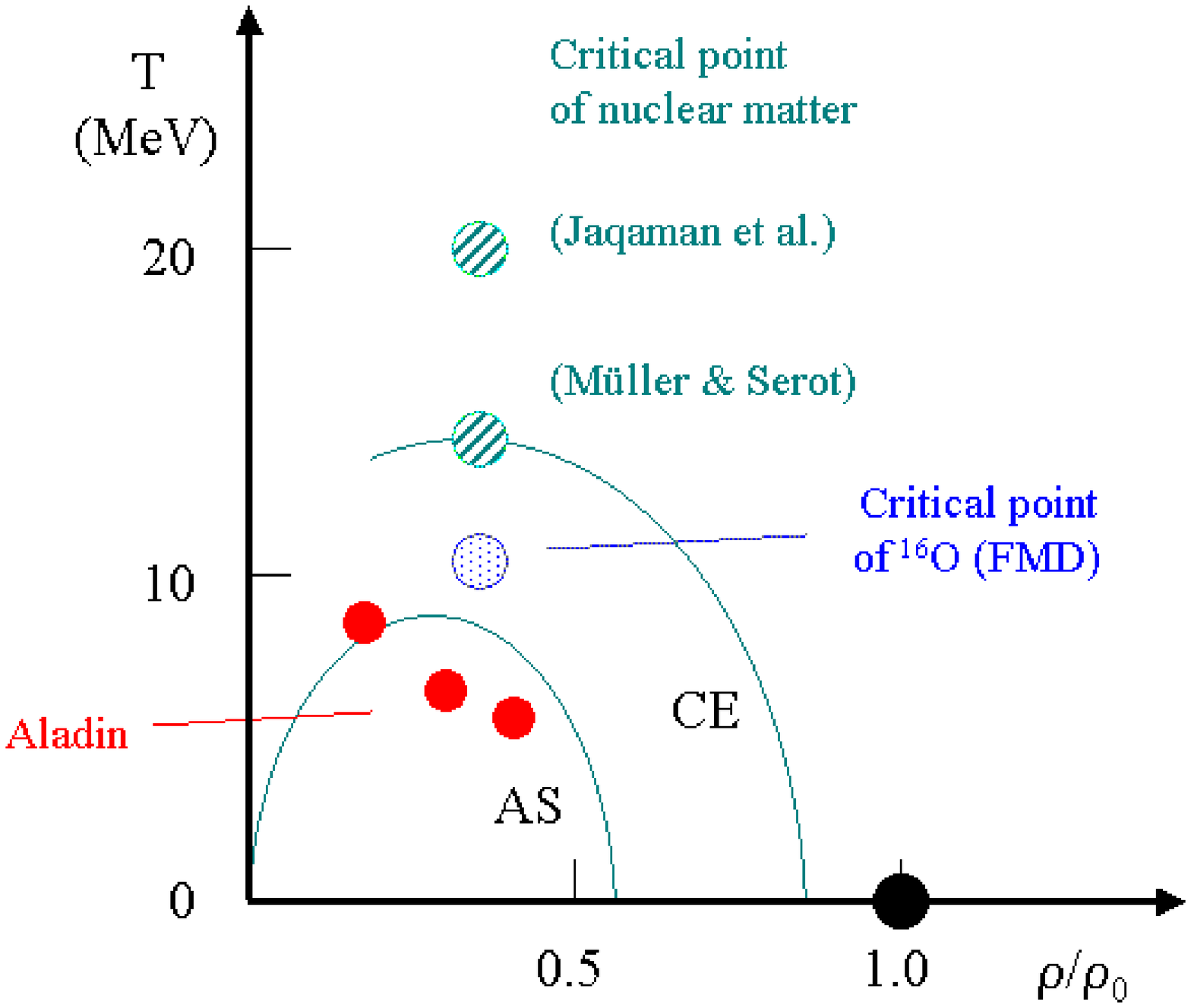}
\end{minipage}
\begin{minipage}[c]{.49\textwidth}
   \centerline{\includegraphics[width=0.95\textwidth]{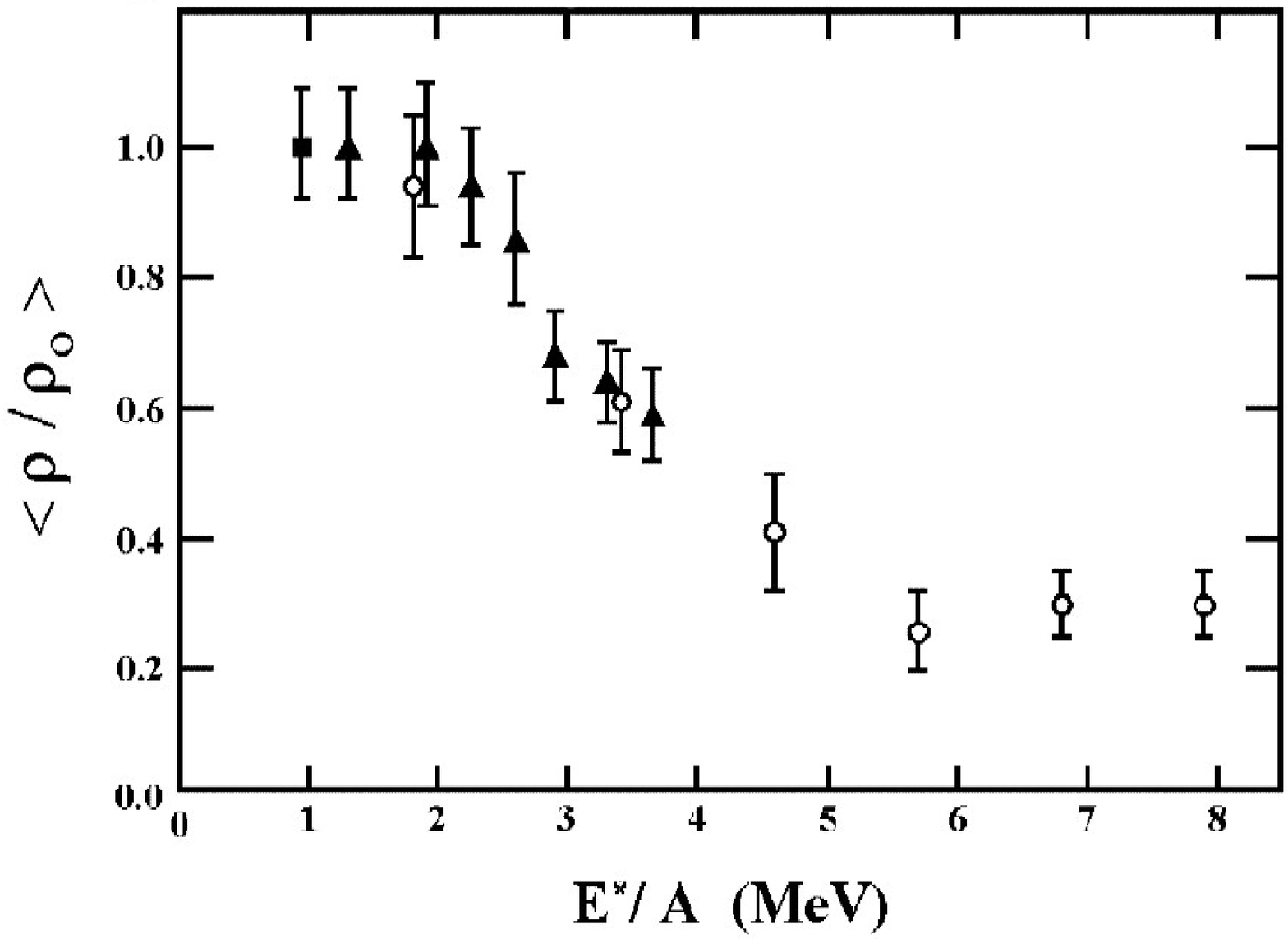}}
\vspace{-5mm}
\caption{Left: temperature-versus-density diagram with the saturation
point of nuclei (closed circle), calculated critical points of nuclear matter
(hatched), the critical temperature of $^{16}$O according to the FMD model
and with experimental results (dots) obtained by correlating measured 
temperatures and densities (from ref.~\protect\cite{nn2000}).
\newline
Top panels: temperatures (left) and densities (right) of fragmenting systems 
at the freeze-out stage as a function of their excitation energy per nucleon.
For details see refs.~\protect\cite{nato02,nn2003,viola04}. 
}
\end{minipage}
\label{phase}
\end{figure}

\begin{figure}[htb]

   \centerline{\includegraphics[height=6.5cm]{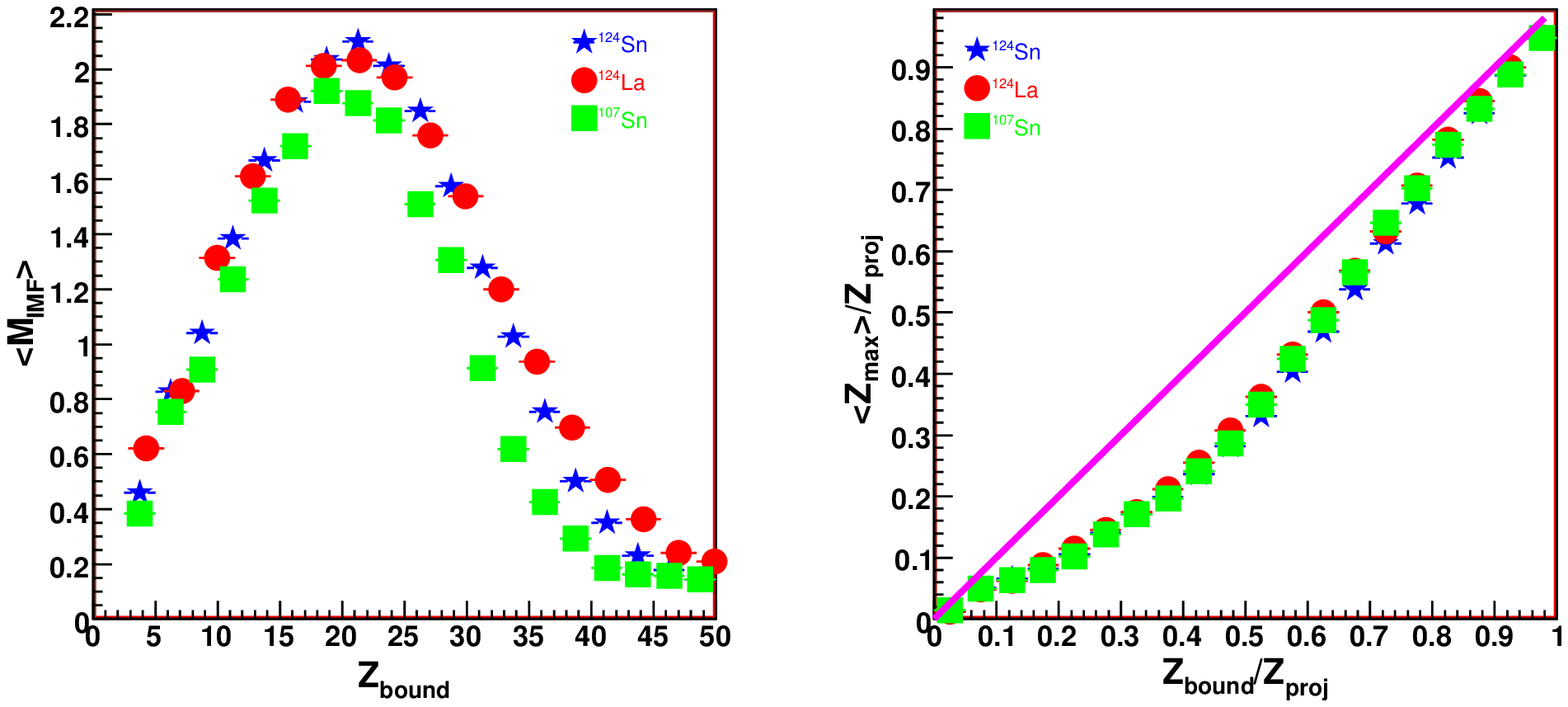}}
\vspace{-4mm}
\caption{Mean multiplicity $<$$M_{\rm IMF}$$>$ of intermediate-mass fragments 
$3 \le Z \le 20$ produced in the fragmentation of $^{107,124}$Sn and $^{124}$La
(600 A MeV, ALADIN) as a function of $Z_{\rm bound}$ (left panel). The right 
panel shows the correlation of the mean $Z$ of the largest fragment with 
$Z_{\rm bound}$ with both quantities normalized with respect to the 
projectile $Z$.
}
\label{global} % optional figure label, must be unique
\end{figure}

\section{GROSS PROPERTIES}

Isotopic effects are rather small for global observables of the studied 
reactions as, e.g., the multiplicity of the produced intermediate-mass fragments
or the correlation of the maximum charge $Z_{\rm max}$ with 
$Z_{\rm bound} = \Sigma Z_i$ with $Z_i \ge 2$ (Fig.~\ref{global}; 
the sorting variable $Z_{\rm bound}$ is related to the impact parameter 
and inversely related with the excitation energy per nucleon of the produced 
spectator system). The universal
rise and fall of fragment production \cite{schuett96} is recovered 
(Fig.~\ref{global}, left panel) and only a slightly steeper slope in the rise
section distinguishes the neutron-rich case of $^{124}$Sn from the other two 
systems. As confirmed by statistical model calculations \cite{sfienti_prag},
this difference is related to the evaporation properties of excited heavy nuclei. 
Neutron emission prevails for neutron-rich nuclei which leads to a concentration 
of the residue channels, with small associated fragment multiplicities, 
in a somewhat narrower range of large $Z_{\rm bound}$ than in
the neutron-poor cases exhibiting stronger charged-particle emissions.
The evaporation of protons reduces $Z_{\rm bound}$ since 
protons are not counted therein.

The effect is, nevertheless, small and nearly invisible in the correlation of 
$<$$Z_{\rm max}$$>$ with $Z_{\rm bound}$ (Fig.~\ref{global}, right panel). In this 
correlation the transition from predominantly residue production 
to multifragmentation becomes apparent as a reduction of $<$$Z_{\rm max}$$>$ with
\linebreak[4] 
respect to $Z_{\rm bound}$, approximately representing the total charge of the 
decaying system,
which occurs at $Z_{\rm bound}/Z_{\rm proj} =$~0.7 to 0.8. Small differences 
are also observed for the fragment $Z$ spectra~\cite{sfienti_nn2006} and
for deduced quantities as, e.g., the breakup temperature 
obtained from double ratios of isotope yields discussed below.
The range of isotopic compositions of the studied nuclei $N/Z = 1.14$~to 1.48
is not sufficiently large, so that significant variations of the reaction 
mechanism would appear. A shrinking of the coexistence zone in the 
temperature-density plane is expected for neutron-rich matter but larger 
consequences can only be expected for asymmetries
far beyond those presently available for laboratory studies \cite{muell95}. 
This fact, on the other hand, has the important consequence that 
the basic reaction process is the same for all the studied cases, 
a prerequisite for the interpretation of isoscaling and its
relation with the symmetry energy.

\section{ISOTOPIC EFFECTS IN LIGHT FRAGMENT PRODUCTION}

The mass resolution obtained for projectile fragments entering into the 
acceptance of the ALADIN spectrometer is about 3\% for fragments with $Z = 3$
and decreases to 1.5\% for $Z\geq 6$.
Masses are thus individually resolved for fragments with atomic number $Z \leq 10$.
The elements are resolved over the full range of atomic numbers up 
to the projectile $Z$ with a resolution of $\Delta Z \leq 0.2$ obtained with the
TP-MUSIC IV detector \cite{sfienti_prag}. 

\begin{figure}[htb]
   \centerline{\includegraphics[width=0.85\textwidth]{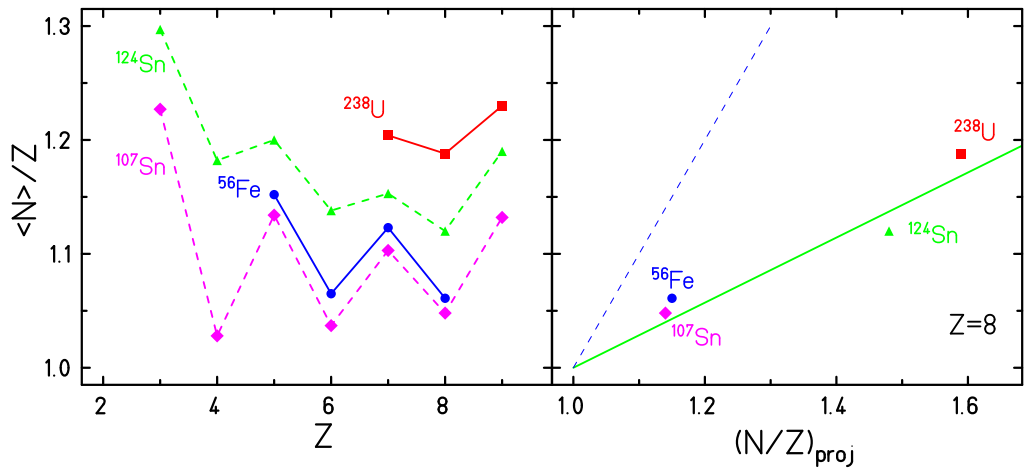}}
\vspace{-4mm}
\caption{Inclusive mean values $<$$N$$>$/$Z$ of light fragments with $3 \le Z \le 9$ 
produced in the fragmentation of $^{107,124}$Sn (600 A MeV, ALADIN),
$^{56}$Fe and $^{238}$U (both 1 A GeV, FRS) as a function  
of the fragment $Z$ (left panel). The right panel shows the results for $Z=8$ 
as a function of the $N/Z$ value of the projectile. The lines represent the trend of the data (full line) and $<$$N$$>$/$Z = (N/Z)_{\rm proj}$ (dashed).
}
\label{noverz} % optional figure label, must be unique
\end{figure}

The mean $N/Z$ of the inclusive isotope distributions of light 
fragments in the range $3 \le Z \le 9$ is presented in Fig.~\ref{noverz}. 
The values for $Z=4$ have been corrected for the missing yield of unstable $^8$Be 
by including an estimate for it obtained from a smooth interpolation over the 
identified 
yields of $^{7,9-11}$Be. This correction has a negligible effect for the case 
of $^{107}$Sn but lowers the $<$$N$$>$/$Z$ of Be for $^{124}$Sn from 1.24 to 1.18 which 
makes the systematic odd-even variation more clearly visible 
for the neutron rich case. 
The odd-even staggering is more strongly pronounced for the neutron-poor $^{107}$Sn.
The strongly bound even-even nuclei attract a large fraction of the product 
yields during the secondary evaporation stage~\cite{ricci04}. Their effect is,
apparently,  
larger if the hot fragments are already close to symmetry, as it is expected 
for the fragmentation of $^{107}$Sn \cite{buyuk05}.

Inclusive data obtained with the FRS fragment separator at GSI for 
$^{238}$U~\cite{ricci04} and $^{56}$Fe~\cite{napo04} fragmentations on titanium 
targets
at 1 A GeV bombarding energy confirm that the observed patterns are very systematic,
exhibiting at the same time nuclear structure effects characteristic for the 
isotopes produced and significant memory effects of the isotopic
composition of the excited 
system by which they are emitted. This has the consequence that, because of its 
strong variation with $Z$, the neutron-to-proton ratio $<$$N$$>$/$Z$ is not a useful 
observable for studying nuclear matter properties. For this purpose,
techniques, such as the isoscaling discussed below, will have to be used which cause the nuclear structure effects to 
cancel out. Selecting a particular element, e.g. $Z=8$, is already sufficient
to reveal the clear correlation of $<$$N$$>$/$Z$ with the $N/Z$ of the projectile 
(Fig.~\ref{noverz}, right panel). According to the Statistical Fragmentation 
Model (SMM, ref.~\cite{bond95}), the correlation should be much stronger for the
hot fragments of the breakup stage and, in fact, not deviate much from the
dashed $<$$N$$>$/$Z = (N/Z)_{\rm proj}$ line in the figure~\cite{buyuk05}. 
The difference is due to sequential
decay and its being directed toward the valley of stability. A precise modeling of
these secondary processes is, therefore, necessary for quantitative analyses.

\begin{figure}[htb]
\centering
\begin{minipage}[c]{.45\textwidth}
   \centerline{\includegraphics[height=8.0cm]{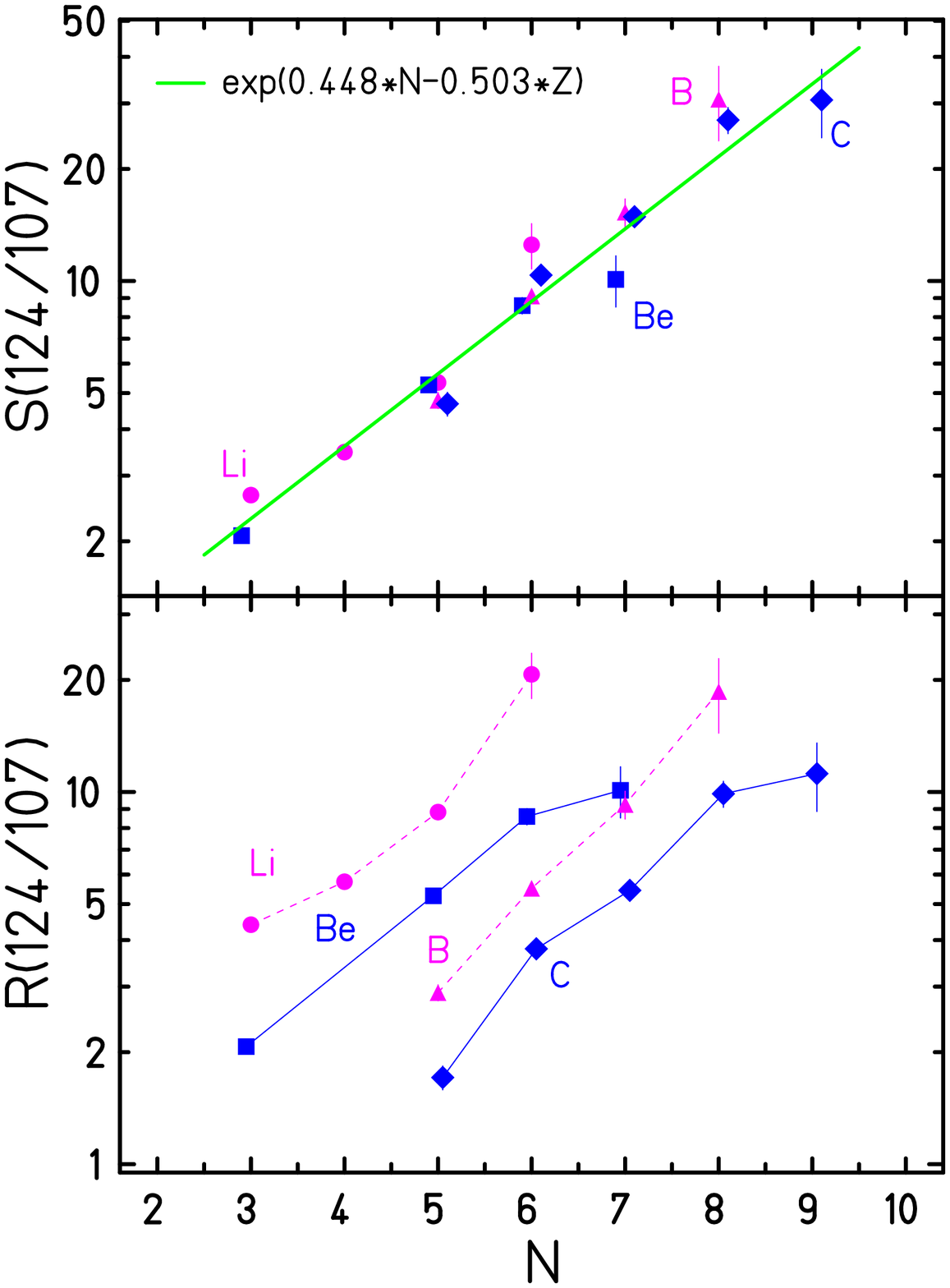}}
\end{minipage}
\begin{minipage}[c]{.53\textwidth}
   \centerline{\includegraphics[height=7.9cm]{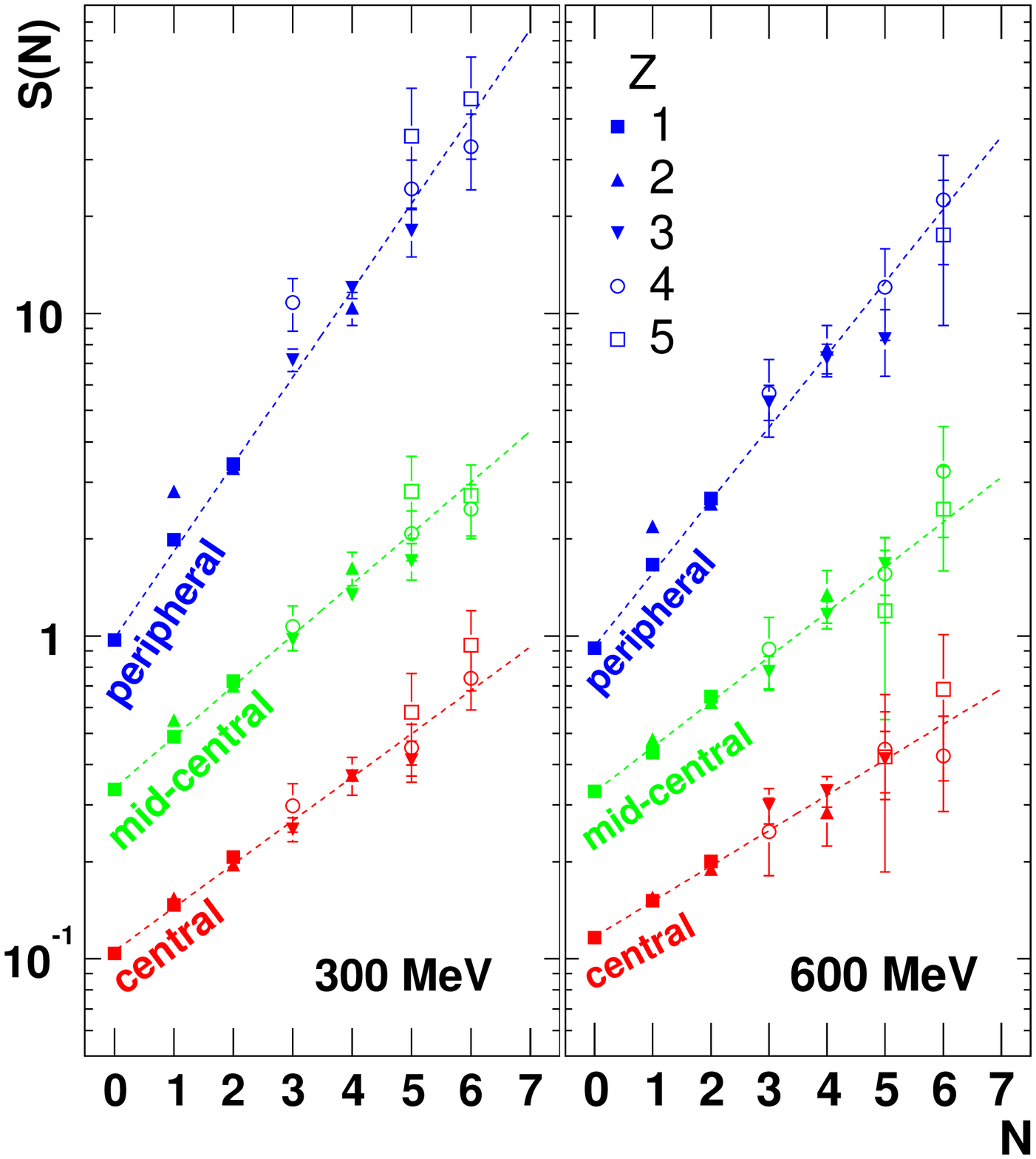}}
\end{minipage}
\vspace{-2mm}
\caption{Left panels: Normalized yield ratios $R(N)$ (bottom) and scaled ratios
$S(N)$ (top) for Li, Be, B, C isotopes from the
fragmentation of $^{124}$Sn and $^{107}$Sn projectiles at 600 A MeV (ALADIN) as a 
function of the neutron number $N$. The full line in the top panel represents the 
exponential fit curve $C'\cdot$exp($\alpha N)$. For clarity, some of the data symbols are slightly displaced horizontally.
\newline
Right panels: Scaled isotopic ratios $S(N)$ for $^{12}$C + $^{112,124}$Sn 
at $E/A$ = 300 and 600~MeV (INDRA) for three intervals of reduced 
impact parameter with
''central'' indicating $b/b_{\rm max} \leq 0.4$ and with offset factors of 
multiples of three. 
H, He, Li, Be and B fragments are distinguished by different data symbols as indicated. 
The dashed lines are the results of exponential fits according 
to Eq.~(\protect\ref{eq:scalab}). Only statistical errors are displayed 
(from ref.~\protect\cite{lef05}).
}
\label{isoscal} % optional figure label, must be unique
\end{figure}

\section{ISOSCALING}

The experimental study of particle and fragment production with 
isotopic resolution
has led to the identification of isoscaling, a phenomenon 
shown to be common to many different types of heavy-ion reactions
\cite{tsang01,botv02,soul03,fried04}.
It is observed by comparing product yields
$Y_i$ from reactions which differ only in the isotopic
composition of the projectiles or targets or both. 
Isoscaling refers to an
exponential dependence of the measured yield ratios $R_{21}(N,Z)$
on the neutron number $N$ and proton number $Z$ of the detected 
products. The scaling expression
\begin{equation}
R_{21}(N,Z) = Y_2(N,Z)/Y_1(N,Z) = C \cdot exp(\alpha N + \beta Z)
\label{eq:scalab}
\end{equation}
describes rather well the measured ratios over a wide range of
complex particles and light fragments \cite{tsang01a}. 
For illustration, the yield ratios $R$(124/107) obtained for the fragmentation 
of the $^{124}$Sn and $^{107}$Sn projectiles in the $Z_{\rm bound} =$~10-30 
interval, covering the region of maximum fragment production, are plotted in 
Fig.~\ref{isoscal} (bottom left). A fairly regular pattern can be recognized,
the yield ratios for a given element increase with $N$ and the lines of different
elements are evenly displaced from each other. Fitting with the expression given
above yields the parameters $\alpha = 0.448$ and $\beta = -0.503$
as indicated in the figure. The quality with which the isoscaling relation is
obeyed is more easily judged 
for the scaled isotopic ratios $S(N) = R_{21}(N,Z)/{\rm exp}(\beta Z)$ from 
which the $Z$ dependence has been removed (Fig.~\ref{isoscal}, top panel).

Within the statistical model, there is a simple physical 
explanation for the appearance of isoscaling in finite systems.
Charge distributions of fragments with fixed mass numbers $A$, as well 
as mass distributions for fixed $Z$, are approximately Gaussian with 
average values and variances which are connected with the temperature,
the symmetry-term coefficient, and other parameters. The mean values depend 
on the total mass and charge of the systems, e.g. via the chemical 
potentials in the grand-canonical approximation, while 
the variances depend mainly on the physical conditions reached,
the temperature, the density and possibly other variables. For example, 
the charge variance $\sigma_Z\approx \sqrt(AT/8\gamma)$ obtained for
fragments with a given mass number $A$ in ref. \cite{botv85} 
is only a function of the temperature and of the symmetry-term coefficient 
$\gamma$~\cite{gamma}
since the Coulomb contribution is very small. This relation of isoscaling
with the symmetry energy has attracted considerable interest recently.

\section{THE SYMMETRY ENERGY}

The dependence of $S(N)$ on the impact parameter for the reactions
$^{12}$C + $^{112,124}$Sn at 300 and 600~MeV per nucleon, 
studied with INDRA at GSI \cite{lef05}, is shown in the right panels of
Fig.~\ref{isoscal}. The slope parameter $\alpha$ decreases strongly 
with increasing centrality to values smaller than 50\%
of those obtained for the peripheral event groups. 
In the grand-canonical approximation,
assuming that the temperature $T$ is about the same,
the scaling parameters $\alpha$ and $\beta$ are given by 
the differences of the neutron and proton chemical potentials for
the two systems divided by the temperature, 
$\alpha = \Delta \mu_{\rm n}/T$ and $\beta = \Delta \mu_{\rm p}/T$.
A proportionality of $\alpha$ and $1/T$, with $T$ derived from double 
isotope ratios, has been observed for light-ion (p, d, $\alpha$)
induced reactions at bombarding energies in the GeV range~\cite{botv02}.
These data were, however, inclusive and thus not representative for 
multi-fragment decays because the mean fragment multiplicities are known to be small
in this case \cite{beaulieu}.

\begin{figure}[htb]
\centering
\begin{minipage}[c]{.49\textwidth}
   \centerline{\includegraphics[height=9.5cm]{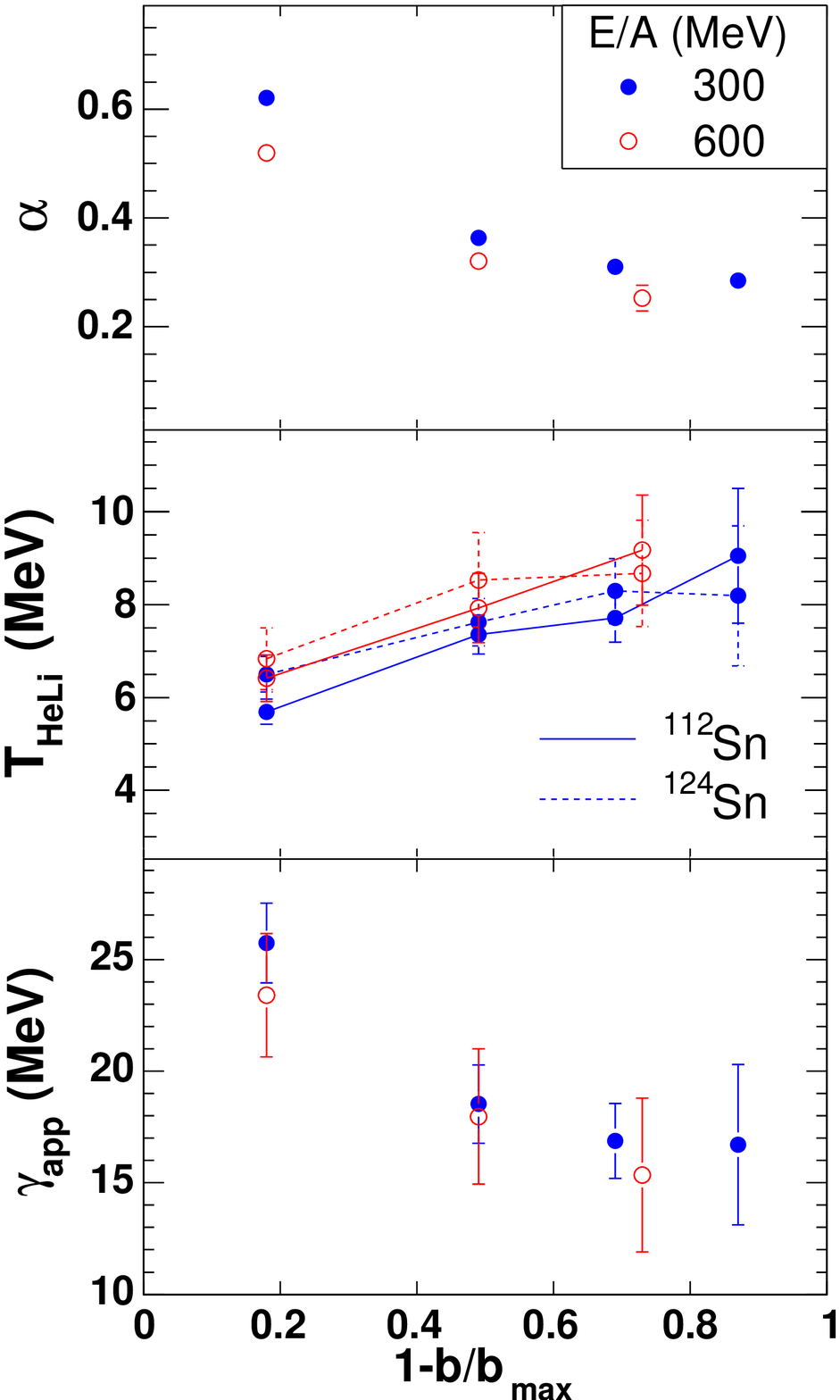}}
\end{minipage}
\begin{minipage}[c]{.49\textwidth}
   \centerline{\includegraphics[height=9.5cm]{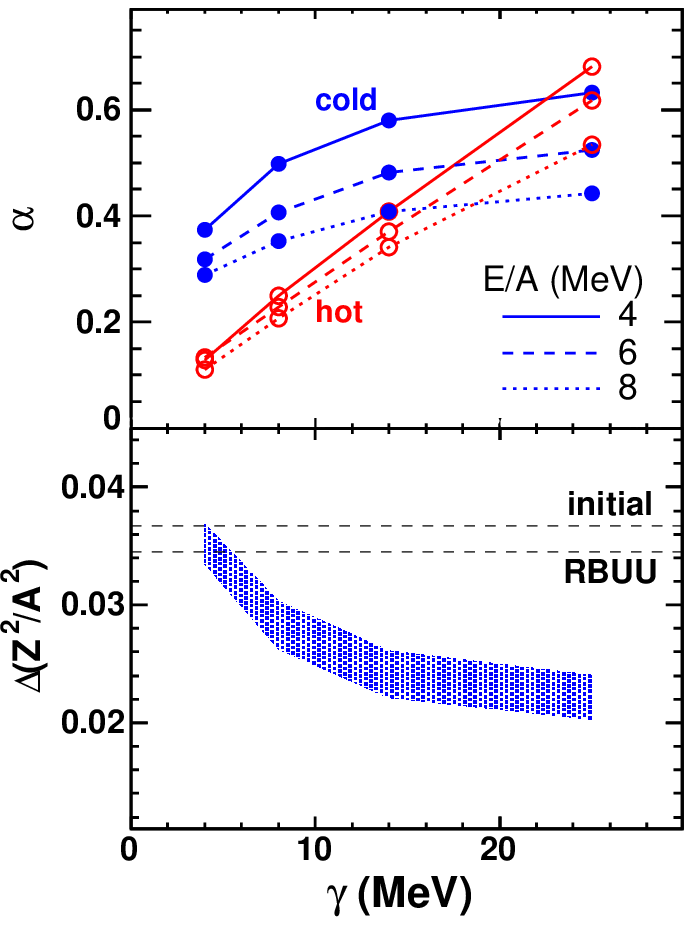}}
\end{minipage}

\caption{The left panels show the 
isoscaling coefficient $\alpha$ (top), the double-isotope temperatures 
$T_{\rm HeLi}$ (middle) and the resulting $\gamma_{\rm app}$ (bottom) 
for the $^{12}$C on $^{112,124}$Sn
reactions at $E/A$ = 300~MeV (full symbols) and 600~MeV (open symbols) 
as a function of the centrality parameter $1-b/b_{\rm max}$.
\newline 
The top right panel shows the isoscaling coefficient $\alpha$ for hot 
(open circles) and cold fragments (dots) as a function of the symmetry-term 
coefficient $\gamma$ as predicted by the Markov-chain calculations for 
$^{112,124}$Sn and excitation energies $E/A = 4,6,8$~MeV.
The shaded area in the bottom-right panel represents the region in the 
$\Delta (Z^2/A^2)$-versus-$\gamma$ plane that is consistent with the measured 
value $\alpha =0.29$ for central collisions
and with the Markov-chain predictions for cold fragments.
The dashed lines indicate the $\Delta (Z^2/A^2) = 0.0367$ of $^{112,124}$Sn 
and the RBUU prediction (from ref.~\protect\cite{lef05}).
}
\label{lefevre} % optional figure label, must be unique
\end{figure}

The $^{12}$C induced reactions cover the rise of fragment production
over the full range up to $Z_{\rm bound} \approx Z_{\rm proj}/2$ with multiple
fragment production being dominant in central collisions~\cite{schuett96}.
The measured temperatures rise slightly with centrality 
(note their similarity for the two
systems) but not sufficiently fast in order to compensate for the decrease of the 
isoscaling parameter $\alpha$ (Fig.~\ref{lefevre}, left panels), and 
$\Delta \mu_{\rm n} = \alpha \cdot T$ decreases according to the 
above relation.

The proportionality of $\Delta \mu_{\rm n}$ and thus of 
the isoscaling parameters with the coefficient $\gamma$ of
the symmetry-energy term $E_{\rm sym} = \gamma (A-2Z)^2/A$
has been obtained from the statistical 
interpretation of isoscaling within the SMM \cite{botv02} and 
Expanding-Emitting-Source Model~\cite{tsang01a} 
and confirmed by an analysis of reaction dynamics
\cite{ono03}. The relation is
\begin{equation} \label{eq:dmunu}
\Delta \mu_{\rm n} = \mu_{\rm n,2} - \mu_{\rm n,1} \approx 4\gamma
(\frac{Z_{1}^2}{A_{1}^2}-\frac{Z_{2}^2}{A_{2}^2}) = 4\gamma \Delta (Z^2/A^2)
\end{equation}
where $Z_{1}$,$A_{1}$ and $Z_{2}$,$A_{2}$ are the charges and mass
numbers of the neutron-poor and neutron-rich systems, respectively, at breakup. 
The difference of the chemical potentials
depends essentially only on the coefficient $\gamma$ of the symmetry term
and on the isotopic compositions. Using this equation and 
assuming that the isotopic compositions are practically
equal to those of the original targets, an apparent symmetry-term 
coefficient $\gamma_{\rm app}$ was determined, i.e. without corrections for the
effects of sequential decay. The results are found to be close to 
the normal-density coefficient $\gamma \approx 25$~MeV for peripheral collisions 
but drop to lower values at central impact parameters (Fig.~\ref{lefevre}, 
bottom left).

The effects of sequential decay for the symmetry term, as calculated with
the microcanonical Markov-chain version of the SMM \cite{botv01}, 
are shown in the top-right panel of Fig.~\ref{lefevre}.
Instead of using the above equations, the isoscaling coefficient $\alpha$ was
determined directly from the calculated fragment yields before (hot fragments) and
after (cold fragments) the sequential-decay part of the code.  
The hot fragments exhibit the linear relation of $\alpha$ with $\gamma$ as expected
but, for $\gamma$ smaller than 25 MeV, the sequential decay causes a narrowing 
of the initially broad isotope distributions which leads to larger $\alpha$ 
for the decay products.
The variation of $\alpha$ with $\gamma$ is thus considerably reduced and 
the value $\alpha < 0.3$ measured for the most central bins is only reproduced 
with input values $\gamma \leq$ 10 MeV, according to the calculations. 
Another possible source of uncertainty is the isotopic composition at breakup.
Transport models predict that their difference for the two systems should not 
deviate by more than a few percent from the original value \cite{gait04} but 
larger deviations would have a significant effect 
as illustrated in Fig.~\ref{lefevre}, bottom right.

\section{DISCUSSION}

A decrease of the isoscaling parameter $\alpha$ with increasing violence
of the collision, beyond that expected from the simultaneous increase of the
temperature, has also been observed for reactions at intermediate energy
(refs.~\cite{shetty05,souliotis06} and references therein) and a systematics 
is emerging. A very recent interpretation of these results arrives at the 
conclusion that it is the reduced density rather than the elevated temperature
which causes the symmetry term to be smaller than the standard 
value~\cite{bao06}. This analysis uses a 
temperature independent potential part of the symmetry term and a kinetic part 
in the form of a Fermi gas. Consequences of the low-density of the considered
homogeneous system thus have to be assumed to be preserved in the process 
of fragment formation.

Statistical multifragmentation models 
as, e.g., the SMM~\cite{bond95} 
consider normal-density fragments 
statistically distributed within an expanded volume. Here a reduction of the
symmetry term can be imagined to be caused by a modification of the fragments 
in the hot environment, including deformations or nuclear 
interactions between them. 
The neglect of such effects in the actual codes should probably be regarded 
as an idealization as the assumption of equilibrium at the final freeze-out stage
requires interactions in order to achieve it. Fragment formation out of an 
expanded system can thus naturally explain variations of the symmetry term. 
The consequences for the predictions of 
this class of models are rather small since the 
partitioning is predominantly driven by the surface term in the liquid-drop description of the fragments while a variation of the symmetry term affects mainly the isotopic
distributions \cite{buyuk05,ogul02,surface06}.

Further studies will be needed in order to quantitatively establish the 
reduction of the symmetry term. The sequential decay corrections are obviously
important but also the evolution of the isotopic compositions during the reaction
process deserves attention. An experimental reconstruction of the
neutron-to-proton ratio of the detected spectator systems seems possible with
the data from the ALADIN experiments. The spectator source of neutrons 
required for this purpose has already been identified with the coincident data 
from the LAND neutron detector~\cite{iwm05}. 

The fruitful collaboration and stimulating discussions 
with my colleagues 
S.~Bianchin, A.S.~Botvina, J.~Brzychczyk, A.~Le F\`evre, J.~{\L}ukasik, 
P.~Pawlowski and C.~Sfienti are 
gratefully acknowledged.


\begin{thebibliography}{9}
\bibitem{gamma}
the notation follows that originally chosen in Ref.~\cite{bond95}; 
alternatively $C_{\rm sym}$ is frequently used for the same quantity.
\end{thebibliography}
\end{document}